\documentclass[lettersize,journal]{IEEEtran}
\IEEEoverridecommandlockouts

\usepackage{cite}
\usepackage{amsmath,amssymb,amsfonts}
\usepackage{textcomp}
\usepackage{stfloats}
\usepackage{booktabs}
\usepackage{array}
\usepackage[table]{xcolor}
\usepackage{graphicx}
\usepackage{url}
\usepackage[colorlinks=true, linkcolor=blue, citecolor=blue, urlcolor=blue, filecolor=blue]{hyperref}
\definecolor{headerblue}{RGB}{31,73,125}
\definecolor{headerfill}{RGB}{232,238,245}
\definecolor{neutralgray}{RGB}{89,89,89}
\definecolor{revgreen}{RGB}{0,120,40}

\definecolor{revblue}{RGB}{0,70,200}

\usepackage{setspace}
\usepackage{balance}

\begin{document}

\title{One Rater to Score Them All: Frozen Large Language Model Anchoring for Video Quality of Experience}

\author{Parsa Hassani Shariat Panahi, Amir H. Jalilvand, and M. Hassan Najafi%
\thanks{P. Hassani Shariat Panahi and A. H. Jalilvand are with the School of Computer
Engineering, Iran University of Science and Technology, Tehran, Iran.%
M. H. Najafi is with the Department of Electrical, Computer and Systems
Engineering, Case Western Reserve University, Cleveland, OH, USA.}}


\maketitle

\begin{abstract}
Quality of Experience results for video streaming cannot be compared across
studies, because public databases share neither test material nor subjects.
This work introduces a \textit{common reference rater}: an open-weights large
language models with a fixed rubric and greedy decoding that scores the
delivery events of every session and supplies the shared items that score
equating requires. The rater is used without training, and it reads only the
playback event log of startup delay, stalls and quality switches, which most
delivery databases retain even when they share nothing else. A monotonic
curve then maps each study's native scale onto the common scale. On 3{,}552
sessions from four studies covering four rating scales, the rater attains a
Spearman correlation of 0.809 on VL04 of the \texttt{ITU-T P.1203} collection,
exceeding \texttt{P.1203} mode~0 on the same sessions, with a Krippendorff alpha of
0.974 over repeated scoring and 0.951 across three serving providers.
Sessions from three further laboratories give correlations from 0.756 to
0.787. After calibration, the four studies are placed on a single 0 to 100
scale with a held-out error between 7.8 and 13.3 points, and a mean VMAF of
60 corresponds to positions 17 points apart in two of the studies.
\end{abstract}

\begin{IEEEkeywords}
Quality of Experience (QoE), large language models  (LLMs),  video streaming, subjective assessment, score calibration.
\end{IEEEkeywords}

\vspace{-1.5em}
\section{Introduction}
\label{sec:introduction}

\IEEEPARstart{Q}{uality}
of Experience (QoE) refers to how a viewer perceives a video stream, and it
governs the design of video delivery systems. The
field measures QoE through subjective experiments, in which viewers rate what
they see, and through objective models that predict those ratings after being
fitted to ratings collected in earlier experiments \cite{itu_p1401}. These
measurements, however, do not accumulate, because each experiment defines its
own conditions and reports on its own scale. Therefore, a mean opinion score
(MOS) of 3.6 on the \texttt{ITU-T P.910} scale, a z-scored rating, and a 0 to
100 panel mean describe different quantities that cannot be converted into
one another. \texttt{ITU-T P.1401} \cite{itu_p1401} aligns a model with
subjective data in a pairwise fashion but discards the fitted mapping once
the evaluation ends. Progress therefore cannot be measured across studies.
The same limitation extends to QoE-based anomaly detection
\cite{qoe_anomaly_survey}. Network-based frameworks that estimate QoE from
network measurements with machine learning face the same problem
\cite{qoe-rf}.

Prior work leaves this problem open. Alignment methods link separate
experiments but require material that public databases do not provide
\cite{pinson2003inlsa,perezortiz2019unified,li2021mixed,alignnet}. Raters
built on large models read viewer comments or encoding metadata, not
delivery events \cite{hassani2025bridging,goring2025llm}.
\texttt{ITU-T P.1401} fits its mapping per experiment rather than carrying it
between them \cite{itu_p1401}. 

\begin{figure}[t]
\centering
\includegraphics[height=0.17\textheight,keepaspectratio]{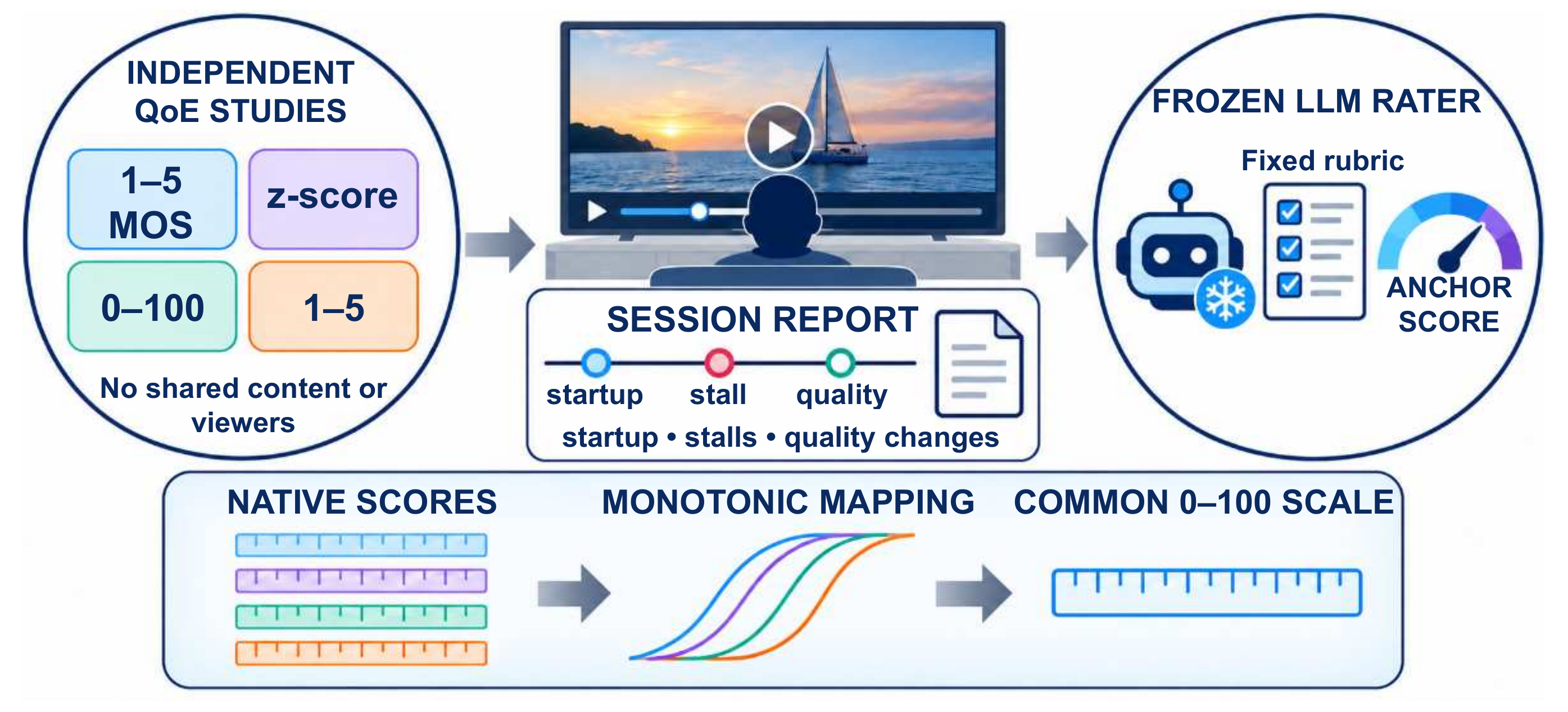}
\vspace{-1.5em}
\caption{{Overview of the proposed procedure. The studies share neither
content nor viewers and report on different scales. Every session is reduced
to a playback log of startup delay, stalls and quality switches, the one
record that all delivery databases retain. A \textit{frozen rater} scores that log
under one fixed rubric, and a monotonic calibration places each study's
ratings on the common 0 to 100 axis.}}
\label{fig:overview}
\vspace{-0.5em}
\end{figure}

In this paper, the studies are bridged by a \textit{common reference rater} rather
than by shared material. Score equating requires common items, and QoE has
none. The rater supplies those items by reading the playback event log of
every session, the one record that all delivery databases retain, and scoring
it under a fixed rubric (Fig.~\ref{fig:overview}). Those scores become the
shared items that score equating requires, and the anchor therefore exists
even when the studies share nothing. This work makes three contributions.
The first is a \textit{reference rater} for delivery QoE that reads structured event
logs as text and requires no access to the decoded video. The second is a
\textit{calibration procedure} that places subjective scores and objective metrics
from independent studies on one common scale, with the residual error
measured out of sample. The third is an evaluation on 3{,}552 sessions from
four studies, including the alternative annotator test
\cite{calderon-etal-2025-alternative} and a reliability characterization over repeated
scoring. The evaluation is
extended by 74 sessions from three further laboratories and by 180 clips
without delivery events, on which the rater is compared with LLMs
that read encoding metadata. The common scale is also tested directly on
sessions that two separate viewer groups rated, and \texttt{ITU-T P.1203} is
compared with the rater in the role of the anchor. 

The rest of the paper is organized as follows.
Section~\ref{sec:related} reviews related work.
Section~\ref{sec:system} describes the proposed rater and its calibration.
Section~\ref{sec:simulation} presents the databases and the evaluation
protocol. Section~\ref{sec:results} reports the results.
Section~\ref{sec:conclusion} concludes the paper.

\section{Related Work}
\label{sec:related}

Prior work on cross-study comparability in QoE falls into three groups. The
first links separate experiments through \textit{shared material}. The second
predicts QoE from delivery events and picture quality. The third builds
raters on large models. The standard
\texttt{ITU-T P.1401} prescribes a mapping fitted per experiment and then
discarded, and is discussed with the first group. Table~\ref{tab:related}
compares representative approaches.

\subsection{Alignment of Separate Experiments}
\label{sec:related_alignment}

Alignment methods link experiments through something they share. Iterated
nested least squares uses an objective video quality metric as the shared
variables, and offers a meta-test that shows selected clips from each dataset
to one panel \cite{pinson2003inlsa}. Unified scaling combines ratings with
pairwise comparisons, and requires that no condition stay disconnected
\cite{perezortiz2019unified}. Mixed-dataset training puts a per-dataset scale
and shift inside one quality model \cite{li2021mixed}. AlignNet fixes one
dataset as the reference and maps the others to it \cite{alignnet}.
Each method needs material that public delivery databases do not have.
Iterated nested least squares needs an objective metric on the decoded video
of every clip \cite{pinson2003inlsa}. Unified scaling needs pairwise
comparisons that cover the datasets \cite{perezortiz2019unified}.
Mixed-dataset training and AlignNet keep the link inside one trained network,
which must be retrained whenever a database is added
\cite{li2021mixed,alignnet}. None of the four puts two studies on a common
scale without that material.
\texttt{ITU-T P.1401} prescribes a mapping between a model and subjective
data \cite{itu_p1401}. The mapping is fitted pairwise, and the evaluation
throws it away once the evaluation ends. A subjective experiment is a closed
set that cannot be compared with other experiments, and the mapping is fitted
per experiment, not carried between them. \texttt{ITU-T P.1401} therefore
does not give the link that cross-study comparison needs.

\begin{table}[t]
\centering
\caption{{Comparison with approaches that link studies. N/A marks an approach
that provides no such link.}}
\label{tab:related}
\renewcommand{\arraystretch}{1.2}
\setlength{\tabcolsep}{3pt}
\scriptsize
\begin{tabular}{@{}
    >{\raggedright\arraybackslash}m{2.3cm}
    >{\centering\arraybackslash}m{0.75cm}
    >{\raggedright\arraybackslash}m{2.75cm}
    >{\raggedright\arraybackslash}m{2.4cm}
    @{}}
\toprule
\rowcolor{headerblue}
{\color{white}\textbf{Approach}} &
{\color{white}\textbf{Year}} &
{\color{white}\textbf{Link between studies}} &
{\color{white}\textbf{What it needs}} \\
\midrule
INLSA \cite{pinson2003inlsa} & 2003 & Objective metric on every clip & Decoded video \\
P.1203 \cite{robitza2018p1203} & 2017 & N/A & Bitstream metadata \\
Unified scale \cite{perezortiz2019unified} & 2020 & Comparisons across datasets & A new subjective experiment \\
P.1401 \cite{itu_p1401} & 2020 & Mapping per experiment & Subjective data \\
Joint training \cite{li2021mixed,alignnet} & 2021--24 & Jointly trained network & Scores of every dataset; retraining \\
LLM on metadata \cite{goring2025llm} & 2025 & N/A & Codec, bitrate, resolution \\
\textbf{This work} &  & \textbf{\texttt{frozen rater}, fixed rubric} & \textbf{The event log alone} \\
\bottomrule
\end{tabular}
\vspace{-1.5em}
\end{table}

\subsection{Objective QoE Models}
\label{sec:related_objective}

Objective QoE models predict the opinion score of a session from its
delivery events and a measure of picture quality, from the bitrate to a
metric computed on the decoded frames \cite{duanmu2022bsqi}. \texttt{ITU-T P.1203}
standardizes a bitstream-based model with four levels of input detail, and
it came with open databases and a reference implementation
\cite{robitza2018p1203}.
A comparison of 15 such models on four databases \cite{duanmu2022bsqi}
reports that learning-based models do well on some test sets and worse on
others, and the Spearman correlation of one model drops by 0.597 between two
databases. Each model is judged by its correlation inside one database, which
does not put the databases on a \textit{common scale}.

\subsection{Raters Built on Large Models}
\label{sec:related_raters}

LLMs have also been used as raters. An LLM has assigned an opinion
score from 1 to 5 to each viewer comment on live-streaming platforms, grouped
per simulated service provider \cite{hassani2025bridging}. LLMs
have also predicted the opinion score of encoded clips from their codec, bitrate
and resolution \cite{goring2025llm}.
Each rater reads a different input. The
comment-based model reads viewer text \cite{hassani2025bridging}, and the
metadata-based models read encoding parameters \cite{goring2025llm}. None reads the \textit{delivery
events} of a session.

\section{Proposed Reference Rater}
\label{sec:system}
The proposed system converts every recorded streaming session into one fixed
\textit{textual report}, scores that report with a frozen LLM, and uses the
resulting scores as the anchor items of a score equating procedure
(Fig.~\ref{fig:overview}). The left branch carries the raw material of a
study through the report format to the rater, which produces an anchor score
on a 0 to 100 scale; the right branch carries the scores the study itself
reports into a monotonic calibration that maps them onto the same axis. The
rater is the only element common to all studies, and therefore the only
element that can carry a scale between them.
The procedure has four steps. The first converts each session into a fixed
report format (Section~\ref{sec:report}). The second describes the rater and
the rubric it follows (Section~\ref{sec:instrument}). The third places each
study's native scale on the common axis (Section~\ref{sec:calibration}). The
fourth states the conditions under which a study can join the common scale
(Section~\ref{sec:protocol}).

\subsection{Session Report}
\label{sec:report}

Delivery QoE depends mainly on startup delay, stalls, and switches of
bitrate and resolution, and the report is restricted accordingly. Each
session is rendered with a fixed field order: viewing device, screen
resolution, video length, startup delay, a timeline of playback events in
wall-clock seconds, and a summary of stall count, total stalled time and
number of quality changes. A playback line is emitted only when the quality
changes and only after a stall. Missing fields are written as \textit{not
reported}, which declares missing information rather than imputing it. The
report carries no study name, identifier or score, which prevents the rater
from recognizing the source of a session.

\subsection{Rating Instrument}
\label{sec:instrument}

An open-weights LLM serves as the rater, prompted with a rubric
that defines six bands on a 0 to 100 scale, each band described by the
delivery conditions that place a session within it. The rubric ranks the
delivery conditions in a fixed order: stalls first, startup delay second,
picture quality relative to the screen third, and the number of quality
changes last. It also declares that a field marked as not reported is
unknown, and states the bitrate a resolution requires in order to appear
clean, because the model otherwise treats a picture that matches the screen
resolution as top quality regardless of its bitrate. Decoding is greedy, with
temperature and nucleus parameters fixed, and each session is submitted in
its own request.

The instrument consists of four archived components: 1) the model weights, 2)
the report format, 3) the rubric text, and 4) the decoding parameters. Any
change to one of them defines a new instrument and requires every session to
be scored again.

\subsection{Calibration onto the Common Scale}
\label{sec:calibration}

Each session has two numbers: the rater score and the value the study reports
for it, which may be a subjective mean opinion score or an objective metric.
For each study, a monotonic logistic curve maps the reported value onto the
rater score. The curve is fitted by least squares, and the study is then
represented on the common scale by the fitted values. The curve is fitted on
one half of the sessions and evaluated on the other, and the halves are
exchanged, so that no session is placed by a curve fitted to itself. One
curve is fitted per study, per viewing context, and per reported metric, so
the same metric can land at different points on the common scale in different
studies.

The logistic curve has four free parameters, and on a small study it can
collapse into a step that assigns nearly every session one of two values. A
study that provides fewer than 60 sessions to a fit is therefore mapped by a
straight line, limited to the range from 0 to 100. From 60 sessions upward
the logistic curve is used. 

\subsection{Submission Protocol}
\label{sec:protocol}

A study joins the common scale under two conditions. It supplies the raw
event logs or, where available, the stimuli of its sessions rather than its
reported scores alone, because reported scores mix method quality with test
difficulty and protocol. Absent fields are labeled. This labeling yields an
explicit coverage label and prevents an incomplete report from being read as
a favorable one. The rater version is then recorded with the resulting
placement.

\section{Simulation}
\label{sec:simulation}

\begin{figure}[t]
\centering
\includegraphics[width=\columnwidth,trim=0cm 0cm 0cm 0.25cm,clip]{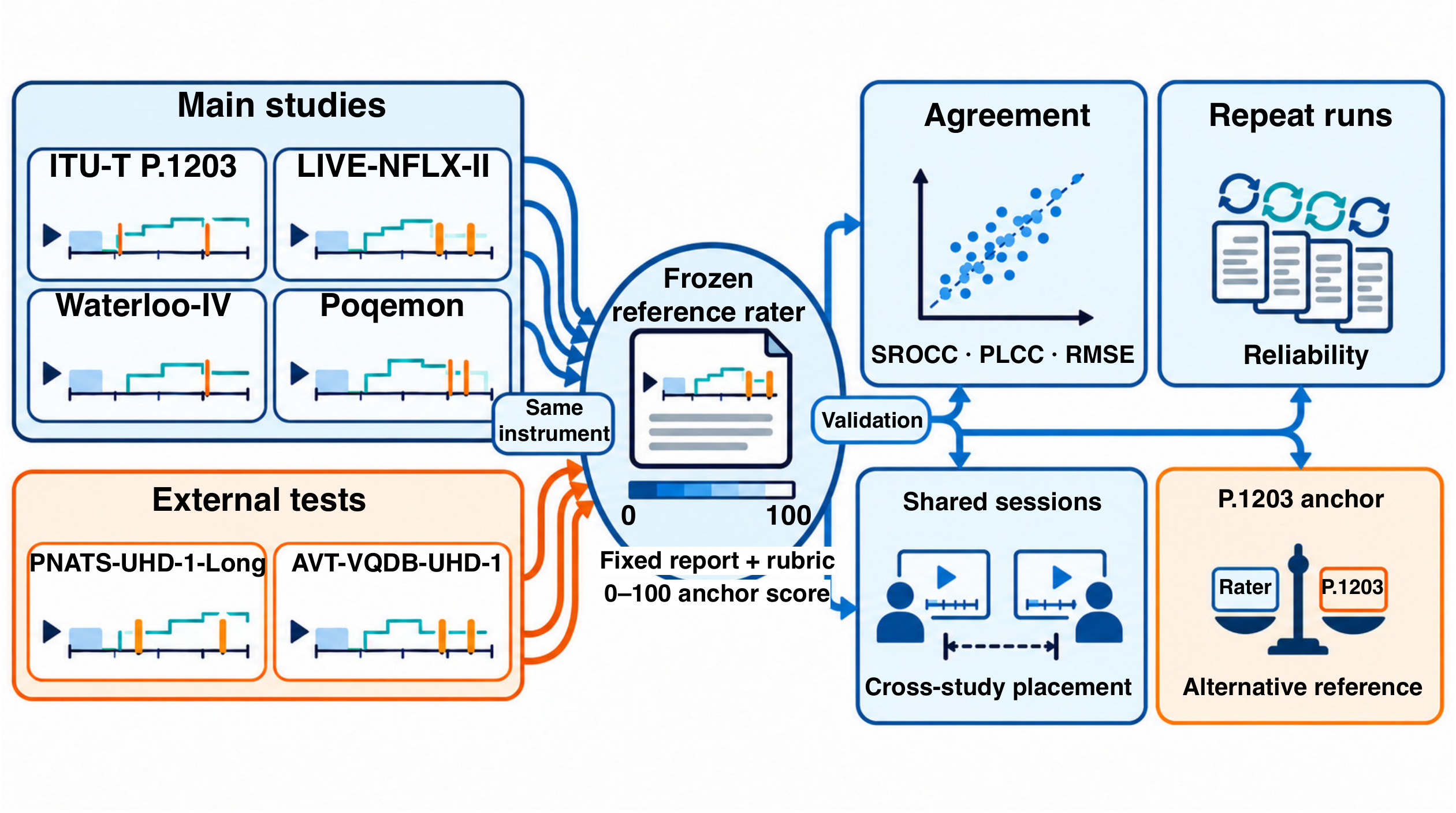}
\vspace{-2.75em}
\caption{{Evaluation setup. The four main studies and the two external test
sets are scored by the same frozen rater. Its scores are used for four tests:
agreement with subjective scores, repeated runs, placement of sessions shared
by two viewer groups, and \texttt{ITU-T P.1203} as an alternative anchor.}}
\label{fig:setup}
\vspace{-1.5em}
\end{figure}

The system is exercised on four independent subjective studies. They share
neither test material nor viewers, and their reported scores lie on four
different scales. The \texttt{ITU-T P.1203} open dataset
\cite{robitza2018p1203} provides 239 sessions rated 1 to 5 by a panel on
scripted event lists. LIVE-NFLX-II \cite{bampis2021towards} provides 420
sessions rated as z-scores under four adaptation algorithms. Waterloo-IV\footnote{\url{https://arxiv.org/abs/2008.08804}} provides 1{,}350 sessions rated 0 to 100 by a panel of 32
under five adaptation algorithms. Poqemon\footnote{\url{https://github.com/Lamyne/Poqemon-QoE-Dataset}} provides 1{,}543
field sessions rated 1 to 5 by a single viewer, reported as totals without a
timeline. The P.1203 dataset contributes five experiments: TR04 and TR06 are
training databases, and VL04, VL13 and VL14 are validation databases; 14
sessions of VL14 carry a subjective score but no feature record and are
excluded. Together the four studies provide 3{,}552 sessions.
Fig.~\ref{fig:setup} summarizes the setup: the four studies and two external
test sets are scored by the same instrument, and its scores feed four tests.

PNATS-UHD-1-Long \cite{rao2023pnats} serves as an external validation set and
was scored only after the instrument had been frozen. It contains three tests
from three laboratories. Rohde \& Schwarz contributed 30 sessions viewed on a
5.5-inch phone, TU Ilmenau and Deutsche Telekom 30 sessions on a 35-inch 4K PC
monitor, and Ericsson Research 14 sessions on a 5.5-inch phone. Sessions last
120 seconds in the first two tests and 300 seconds in the third, and panels of
27 to 32 viewers rate them from 1 to 5.

Test~1 of AVT-VQDB-UHD-1 \cite{rao2019avt} serves as a second external set and
is used for a comparison with LLMs that read encoding metadata
\cite{goring2025llm}. It holds 180 clips of 10 seconds from six sources,
encoded with H.264, H.265 and VP9 at 360p to 2160p and at 200 kbit/s to
40 Mbit/s, and rated by 26 viewers. The clips contain no stalls and no quality
changes. Their metadata and opinion scores are taken from the public data of
\cite{goring2025llm}. The report states the resolution, the bitrate and a
screen of 3840 by 2160 pixels; codec and frame rate have no field in the
report format.

Every session is converted to the report of Section~\ref{sec:report} and
scored once by the reference rater, Qwen3-235B-A22B-2507, served through a
commercial inference provider with temperature 0, a nucleus sampling threshold
of 1, and one session per request. Instrument selection, namely the choice of
model, the output scale and the rubric wording, was performed on TR04 and TR06
only; the validation databases were not consulted at any point.

Three baselines are used. \texttt{ITU-T P.1203} in modes 0 through 3 is
available for the \texttt{P.1203} study. Mean VMAF, PSNR and SSIM are available for
LIVE-NFLX-II, and mean VMAF for Waterloo-IV. A second \texttt{frozen rater},
DeepSeek-V3.2-exp under the identical report format and rubric, is scored on
all four studies to evaluate a panel of raters. In addition, the comparison of
15 objective QoE models on LIVE-NFLX-II and Waterloo-IV published in
\cite{duanmu2022bsqi} is used. There, the models were optimized on two other
Waterloo databases, except \texttt{P.1203}, whose training is not specified, and four
continuous-time models, which were trained on LIVE-NFLX-I. The correlations of
that work are computed on the same 420 and 1{,}350 sessions, and only the
Spearman correlation is compared, because it does not depend on a mapping.

Agreement with subjective scores follows \texttt{ITU-T P.1401}
\cite{itu_p1401}: a monotonic logistic curve is fitted between the predicted
and the subjective values, and the Pearson correlation and the root mean
square error are computed after that mapping, while the Spearman correlation
is computed on the unmapped values. Reliability is characterized by five
repeated scorings of the \texttt{P.1203} study under a pinned provider, by scoring the
same sessions through three different providers, and by a rubric
perturbation. Replacement of a human viewer is tested with the alternative
annotator test \cite{calderon-etal-2025-alternative}, which compares the rater
against each of the 28 viewers of the \texttt{P.1203} study in turn, corrects the
resulting p-values with the Benjamini-Yekutieli procedure, and reports the
winning rate, the share of viewers the rater outperforms at a given tolerance
margin, together with the average advantage probability.

The direct test of the common scale requires sessions that two separate groups
of viewers rated. Waterloo-IV assigned each viewer at random to one of three
devices, a phone, a Full-HD monitor and a UHD television,
and each group rated the same 450 conditions. Since every group had its own
playback run, a condition is retained only where the playback is identical in
both groups, which leaves 283 to 308 shared sessions per pair of groups. Each
group's curve is fitted only on its sessions that are not shared, the shared
sessions are placed by both curves, and the mean absolute gap between the two
placements is reported. The same test is applied to TR04 and VL04, which
contain three sessions with the same source and condition identifiers, and to
the first two tests of PNATS-UHD-1-Long, which share four anchor conditions. In
the \texttt{P.1203} collection, the 82 conditions of TR04 and TR06 are also rated by a
PC panel and by a phone panel. Every condition is shared there, so the
half-split fit of Section~\ref{sec:calibration} is used. The gap expected from
the finite panels alone is obtained from the standard deviation of the
individual ratings of each session and serves as a noise floor.

To determine whether a standardized model could serve as the anchor instead,
\texttt{P.1203} mode~0 is computed for every session whose source data
provide the inputs the model requires. The reference implementation
\cite{robitza2018p1203} is used, and sessions encoded with H.265/HEVC or VP9
use a published codec
extension\footnote{\url{https://github.com/Telecommunication-Telemedia-Assessment/itu-p1203-codecextension}},
which is not part of the standard. The output on the 1 to 5 scale is stretched
linearly onto 0 to 100, and the shared-session test of Waterloo-IV is repeated
with \texttt{P.1203} as the anchor.

Two further experiments concern the design of the instrument. The size below
which the logistic curve becomes unreliable is determined by subsampling. For
each sample size from 8 to 100 sessions, a sample is drawn from a large study
60 times, and the curve fitted on them is evaluated on the remaining sessions.
A second rubric judges sharpness by the resolution of the picture and the type
of device, rather than by the ratio between the picture and the screen
resolution. Its decision rule was fixed before it was scored. The second
rubric would replace the original one only if no study outside its design set
lost more than 0.02 in Spearman correlation, and only if the device offsets of
Waterloo-IV decreased.

\section{Results}
\label{sec:results}

The evaluation answers four questions: how closely the reference rater agrees
with subjective scores (Section~\ref{sec:res_accuracy}), how it compares with
the standardized model (Section~\ref{sec:res_baseline}), how stable the
instrument is under repetition and relocation
(Section~\ref{sec:res_reliability}), and whether the calibrated common scale
renders independent studies comparable (Section~\ref{sec:res_scale}).
Table~\ref{tab:results} reports agreement, coverage and placement error for
every study, and the four subsections take the questions in turn.

\subsection{Agreement with Subjective Scores}
\label{sec:res_accuracy}

Table~\ref{tab:results} reports agreement for each study. The rater attains
Pearson correlations between 0.533 and 0.710 and Spearman correlations
between 0.397 and 0.728 across four scales and four protocols, without access
to the decoded video and without training. Agreement is highest on
LIVE-NFLX-II, whose sessions follow real traces, and lowest on Poqemon, whose
reports carry no timeline and whose scores come from one viewer rather than a
panel.

\begin{table*}[t]
\centering
\caption{{Agreement with subjective scores and coverage of the two candidate
anchors. Pearson correlation (PLCC) and RMSE follow the mapping of
\texttt{ITU-T P.1401}~\cite{itu_p1401}; RMSE is on each study's native scale.
\texttt{ITU-T P.1203} mode~0~\cite{robitza2018p1203} is evaluated on the sessions it
can score. Error is the held-out placement error on the common scale, in
points. The last four rows are external tests, scored after the instrument was
frozen. Empty cells were not computed for that row.}}
\label{tab:results}
\renewcommand{\arraystretch}{1.2}
\setlength{\tabcolsep}{4pt}
\footnotesize
\begin{tabular}{@{}%
    >{\raggedright\arraybackslash}m{3.4cm}%
    >{\centering\arraybackslash}m{1.1cm}%
    >{\centering\arraybackslash}m{1.3cm}%
    >{\centering\arraybackslash}m{1.3cm}%
    >{\centering\arraybackslash}m{1.3cm}%
    >{\centering\arraybackslash}m{1.6cm}%
    >{\centering\arraybackslash}m{1.3cm}%
    >{\centering\arraybackslash}m{1.3cm}%
    >{\centering\arraybackslash}m{1.3cm}%
    @{}}
\toprule
\rowcolor{headerblue}
{\color{white}\textbf{Study}} &
{\color{white}\textbf{Sess.}} &
{\color{white}\textbf{Rater PLCC}} &
{\color{white}\textbf{Rater SROCC}} &
{\color{white}\textbf{Rater RMSE}} &
{\color{white}\textbf{Scored by P.1203}} &
{\color{white}\textbf{P.1203 PLCC}} &
{\color{white}\textbf{P.1203 SROCC}} &
{\color{white}\textbf{Error (points)}} \\
\midrule
P.1203 TR04 (train) & 120 & 0.674 & 0.652 & & 120 & 0.888 & 0.847 & \\
P.1203 TR06 (train) & 44 & 0.711 & 0.648 & & 44 & 0.929 & 0.900 & \\
\textbf{P.1203 VL04 (valid.)} & 60 & \textbf{0.818} & \textbf{0.809} & & 60 & 0.780 & 0.754 & \\
P.1203 VL13 (valid.) & 15 & 0.822 & 0.817 & & 15 & 0.902 & 0.854 & \\
ITU-T P.1203 \cite{robitza2018p1203} & 239 & 0.691 & 0.631 & 0.70 & 239 & & 0.837 & 13.3 \\
LIVE-NFLX-II \cite{bampis2021towards} & 420 & 0.710 & 0.728 & 0.58 & 420 & & 0.833 & 7.8 \\
Waterloo-IV & 1{,}350 & 0.586 & 0.561 & 12.95 & 1{,}350 & & 0.648 & 11.3 \\
Poqemon & 1{,}543 & 0.533 & 0.397 & 0.89 & 0 & -- & -- & 8.9 \\
\midrule
PNATS test 1 \cite{rao2023pnats} & 30 & 0.827 & 0.756 & 0.57 & 26 & & 0.836 & \\
PNATS test 2 \cite{rao2023pnats} & 30 & 0.789 & 0.775 & 0.58 & 26 & & 0.807 & \\
PNATS test 3 \cite{rao2023pnats} & 14 & 0.861 & 0.787 & 0.51 & 12 & & 0.515 & \\
AVT-VQDB-UHD-1 \cite{rao2019avt} & 180 & 0.843 & 0.813 & 0.60 & 180 & & 0.843 & \\
\bottomrule
\end{tabular}
\vspace{-1em}
\end{table*}

The three external tests of PNATS-UHD-1-Long exceed every study of the main
corpus in Spearman correlation. The tests are small, and the 95\% bootstrap
intervals are correspondingly wide: from 0.50 to 0.89, from 0.56 to 0.91 and
from 0.39 to 0.95. AVT-VQDB-UHD-1, whose clips differ only in encoding,
attains the highest Spearman correlation of the rater in
Table~\ref{tab:results}.

A panel of two raters was also evaluated on the three panel-rated studies, by
averaging the ranks of the reference rater and of DeepSeek-V3.2-exp under the
identical report and rubric. Agreement improved on Waterloo-IV (0.586 to
0.599) and decreased on the other two, where the second rater attained only
0.598 and 0.486. Aggregation is therefore beneficial only among members of
comparable strength.

Among the 15 models evaluated in \cite{duanmu2022bsqi}, the rater occupies a
middle position, with Spearman correlations of 0.728 on LIVE-NFLX-II and 0.561
on Waterloo-IV against 0.906 and 0.699 for the best published models. On each
of the two databases the rater exceeds nine of the models and is exceeded by
six. SQI, BSQI, Bentaleb2016 and \texttt{P.1203} exceed it on both, Xue2014 and Liu2012
only on LIVE-NFLX-II, and Video ATLAS and Spiteri2016 only on Waterloo-IV. The
rater was neither trained nor tuned on either database, while most of the
published models were optimized on other subjective databases.

The published Spearman correlations~\cite{duanmu2022bsqi}, on LIVE-NFLX-II and
on Waterloo-IV respectively, are 0.516 and 0.056 for Mok2011, 0.549 and 0.082
for FTW, 0.433 and 0.132 for NARX-QoE, 0.778 and 0.219 for Xue2014, 0.685 and
0.255 for Bi-LSTM, 0.710 and 0.299 for LSTM-QoE, 0.635 and 0.395 for TV-QoE,
0.732 and 0.468 for Liu2012, 0.686 and 0.541 for Yin2015, 0.673 and 0.670 for
Video ATLAS, 0.711 and 0.662 for Spiteri2016, 0.883 and 0.692 for Bentaleb2016,
0.906 and 0.690 for SQI, and 0.893 and 0.699 for BSQI. That comparison reports
0.821 and 0.668 for P.1203 mode~0. Computed on the sessions of this work,
mode~0 attains 0.833 and 0.648 (Table~\ref{tab:results}).

The LLMs of \cite{goring2025llm} were evaluated on the 180 clips of
AVT-VQDB-UHD-1, and their released predictions permit a comparison on
identical clips. Following that work, the correlations are computed on
unmapped scores and the error on the 1 to 5 scale, to which the rater's score
is converted linearly. The rater attains a Pearson correlation of 0.823, and
among the 17 single models of that work it would rank third. The best single
model, DeepSeek-V3 used through its web interface, attains 0.870, a difference
with a 95\% paired bootstrap interval from 0.001 to 0.104. The best group of
eight models attains 0.902, an interval from 0.041 to 0.126.

In Pearson, Kendall and Spearman correlation and in error on the 1 to 5 scale,
the rater attains 0.823, 0.657, 0.813 and 0.698. The corresponding values are
0.902, 0.720, 0.897 and 0.523 for the best group of eight models, 0.895, 0.706,
0.887 and 0.534 for the mean of 11 models, 0.870, 0.734, 0.877 and 0.569 for
DeepSeek-V3 used through its web interface, 0.857, 0.710, 0.860 and 0.586 for
Gemini-2.0-flash-light, and 0.810, 0.673, 0.826 and 0.681 for
Gemini-2.0-flash~\cite{goring2025llm}. The rater therefore exceeds
Gemini-2.0-flash in Pearson correlation and trails it in the two rank
correlations.

The conditions of the two sides differ. The prompt of \cite{goring2025llm} was
selected after other prompts had been tried on these clips, and the group of
eight was chosen among all 131{,}072 subsets against the same opinion scores.
The rater was frozen before the clips were seen and receives less information,
as described in Section~\ref{sec:simulation}.

\subsection{Comparison with the Standardized Model}
\label{sec:res_baseline}

\begin{figure}[t]
\centering
\includegraphics[width=\columnwidth,trim=0cm 0cm 0cm 0.3cm,clip]{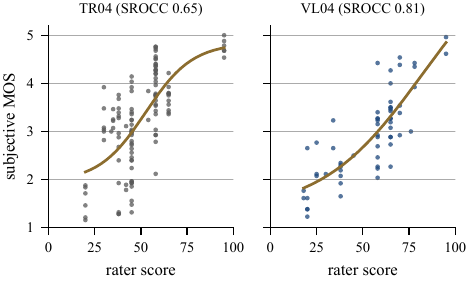}
\vspace{-2.25em}
\caption{{Rater score against subjective mean opinion score for a training
database (left) and a validation database (right), with the fitted monotonic
curve. The Spearman correlation is 0.65 on TR04 and 0.81 on VL04. The wide
residual band on TR04 corresponds to conditions that differ only in encoding
quality at a fixed resolution, which the session report does not express.}}
\label{fig:scatter}
\vspace{-1.5em}
\end{figure}

Table~\ref{tab:results} compares the rater with \texttt{ITU-T P.1203} mode~0 on
identical sessions. On the training databases the standardized model is
clearly superior; on VL04 the ordering reverses, and on VL13 the two are
within 0.05 of each other. What each database varies explains the difference:
TR04 and TR06 contain many conditions that differ only in encoding quality at
a fixed resolution, which a bitstream-level model observes directly and a
textual event report cannot express, whereas VL04 varies resolution, bitrate
and stalling jointly (Fig.~\ref{fig:scatter}).

Replacement of a human viewer was tested with the alternative annotator test
against each of the 28 viewers in turn. On VL04 under PC viewing the null
hypothesis is rejected for 0.88 of the viewers at a tolerance margin of 0.1
and for 1.00 at a margin of 0.2, with an average advantage probability of
0.677; replacement requires a winning rate of at least 0.5. Over the PC
sessions of the whole study the winning rate is 0.50 at a margin of 0.1 and
0.68 at a margin of 0.15. On the training databases the test is not passed, in
agreement with Table~\ref{tab:results}.

Two candidates for the anchor are compared in Table~\ref{tab:results}. Where
\texttt{P.1203} mode~0 can be computed, it agrees more closely with the viewers. The
difference is largest on the \texttt{P.1203} collection, which contains its training
databases, and only on PNATS-UHD-1-Long, pooled over the 64 sessions that both
score, does the rater agree more closely (0.818 against 0.768). However,
\texttt{P.1203} requires inputs that several databases do not report. It scores 2{,}009
of the 3{,}552 sessions of the four studies and none of Poqemon, whose records
carry no per-segment information. A study without anchor scores cannot be
placed on the common scale.

Placement differs as well. Repeated with \texttt{P.1203} as the anchor, the three
Waterloo-IV pairs give gaps of 5.3, 11.6 and 17.8 points, against 5.7, 4.0 and
6.4 points with the rater. Divided by the spread of the respective anchor,
which removes the difference in how the two anchors use the scale, the mean
gap is 0.58 for \texttt{P.1203} and 0.38 for the rater. The standardized model carries
a larger device offset. On the UHD television it scores a stream 14.9 points
below the Full-HD monitor and 18.3 points below the phone. The corresponding
differences of the rater are 5.6 points (Section~\ref{sec:res_scale}) and 9.3
points, and the viewers rated the UHD group higher than both. Agreement within
a study and consistency across studies are therefore separate properties, and
an anchor requires the second.

\subsection{Reliability of the Instrument}
\label{sec:res_reliability}

Five repeated scorings of the 239 sessions under a pinned provider give a
Krippendorff alpha of 0.974 on the interval metric, with a mean standard
deviation of 1.61 points per session against a between-session standard
deviation of 17.1 points. Scoring through three different providers gives an
alpha of 0.951, with mean absolute differences between 1.8 and 3.8 points.
Exact numerical equality is stricter and is not attained: all five repetitions
agree exactly on 53.6 percent of sessions and fall within 10 points on 92.5
percent. The residual variation originates in the serving stack, because
providers batch requests and batching alters the arithmetic; the serving
configuration therefore belongs to the instrument description. A single run on
controlled hardware with batch size one remains to be performed in order to
bound the variation attributable to batching alone.

Rubric perturbation matters as well: adding the paragraph that states the
bitrate each resolution requires altered scores by 4.95 points on average,
approximately three times the repetition noise.

The second rubric of Section~\ref{sec:simulation} addressed the device offset
reported in Section~\ref{sec:res_scale}. It reduced the rater's own gap between
the Full-HD monitor and the UHD television versions of a stream from 5.6 to
1.8 points. The shared-session gap between these two groups fell from 4.0 to
2.4 points, against a panel noise of 2.6 points. It also lowered the Spearman
correlation on LIVE-NFLX-II from 0.728 to 0.611 and on Poqemon from 0.397 to
0.323, while PNATS-UHD-1-Long rose from 0.787 to 0.807. Two studies thus lost
more than 0.02, so by the rule of Section~\ref{sec:simulation} the original
rubric remains the instrument, and the second rubric is reported as an
ablation.

\subsection{Cross-Study Comparability}
\label{sec:res_scale}

\begin{figure}[t]
\centering
\includegraphics[width=\columnwidth]{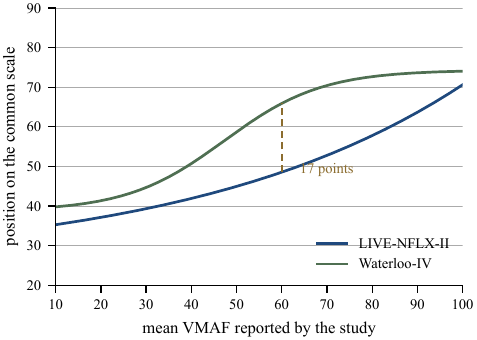}
\vspace{-1.5em}
\caption{{Calibration of mean VMAF onto the common scale for the two studies
that report it. The same metric value occupies different positions in the two
studies, which is the comparability gap the calibration removes.}}
\label{fig:bridge}
\vspace{-1em}
\end{figure}

Measured on held-out sessions, the placement error of the per-study
calibration is 13.3 points for \texttt{ITU-T P.1203}, 7.8 for LIVE-NFLX-II,
11.3 for Waterloo-IV and 8.9 for Poqemon. Pooled over all 3{,}552 sessions the
calibrated subjective scores attain a Pearson correlation of 0.846 with the
anchor scores, against a root mean square error of 10.2 points. Objective
metrics are placed on the same axis by the same procedure. Fig.~\ref{fig:bridge}
shows the curves fitted for mean VMAF in the two studies reporting it. A mean
VMAF of 60 corresponds to position 49 on the common scale in LIVE-NFLX-II and
to 66 in Waterloo-IV. That separation of 17 points quantifies the
comparability gap between two studies reporting the same metric, and the two
curves remove it.

A worked example shows the same effect on subjective scores. In the \texttt{P.1203}
collection, session SRC14\_HRC17 of TR06 plays at 1080p for 105 seconds, drops
to 480p and 240p for 30 seconds, and contains one stall of 15 seconds; its
viewers rated it 3.67 on the 1 to 5 scale. In Waterloo-IV, session 0710 of the
Full-HD monitor group plays at 540p throughout, after a startup delay of 2.1
seconds and one stall of 0.2 seconds; its viewers rated it 39.9 on the 0 to 100
scale. A linear stretch of the two native scales places the first session at
67 and the second at 40. The calibration curves, fitted on all sessions of
each study, place both at 55. The two ratings therefore describe the same
position on the common scale, although their stretched native values differ by
27 points.

The three pairs of Waterloo-IV viewer groups share 308, 288 and 283 sessions:
phone and Full-HD monitor, Full-HD monitor and UHD television, and phone and
UHD television. After calibration, the two placements of a session differ on
average by 5.7, 4.0 and 6.4 points, against a panel noise of 2.7, 2.2 and 2.6
points.

In every pair both groups rated on the same native scale, so their raw scores
were already close; on Waterloo-IV the gaps before calibration are 5.0, 5.6
and 5.4 points for the three pairs. The test therefore measures the precision
of the placement rather than an improvement over a scale the groups already
shared. The calibration is intended for studies on different native scales,
where no raw comparison exists. Its result does not depend on the native
scale: when the phone panel of the \texttt{P.1203} collection is converted to z-scores
before calibration, the gap remains 3.5 points.

Calibration reveals two effects of the anchor. When the rater judges both
versions of a session alike, the calibration removes an offset between panels.
The phone panel rated the 82 shared conditions 6.7 points higher than the PC
panel; after calibration the mean difference is 0.4 points, and the mean
absolute gap falls from 8.5 to 3.5 points, against a panel noise of 2.2 points.
When the rater perceives a device difference that viewers do not, it
introduces an offset of its own. It scores a stream 5.6 points lower on the UHD
television than on the Full-HD monitor, whereas the UHD group rated the same
sessions 5.2 points higher. A mean difference of 3.8 points remains after
calibration.

The two smallest pairs, TR04 with VL04 and the first two PNATS tests, share
only three and four sessions and show gaps of 9.7 and 10.8 points, against 1.6
and 7.4 points before calibration and a panel noise of 1.5 and 1.1 points.
TR04 is dominated by conditions the session report cannot express
(Section~\ref{sec:res_baseline}). The anchor conditions of the first two PNATS
tests are not identical either: their quality ladders differ on two sessions,
and the first test adds an initial stall of 1 second. On the 26 sessions that
the first PNATS test provides to its fit, the logistic curve degenerated, which
motivated the size rule of Section~\ref{sec:calibration}.

Subsampling of LIVE-NFLX-II, Poqemon and the three Waterloo-IV groups yields
the rule. With 26 or fewer sessions, the logistic curve collapsed into a step
in 18 to 55 percent of the draws. Its mean held-out error exceeded 15 points
for every sample of 20 sessions or fewer, whereas a straight line stayed
between 10.3 and 11.5 points. From 60 sessions upward the mean errors of the
two forms differ by at most 0.2 points. A leave-one-out choice between the two
forms performed worse than the straight line alone and was discarded. Under
the rule, the gap of the PNATS pair falls from 10.8 to 5.9 points and that of
the PC and phone panels from 3.5 to 3.2 points. The gap of TR04 and VL04 rises
from 9.7 to 15.8 points, and the Waterloo-IV pairs, fitted on 142 to 167
sessions, are unchanged. The fits behind Section~\ref{sec:res_scale} use at
least 119 sessions each, so the values reported there do not change.

\section{Conclusion}
\label{sec:conclusion}

This work addressed the absence of a common scale in video QoE measurement.
Public delivery databases share neither content nor viewers, and the pairwise
mapping of \texttt{ITU-T P.1401} is discarded once an evaluation ends. The
proposed solution is a \textit{common reference rater}: a frozen open-weights
LLM that scores the textual event report of a session under one
fixed rubric. On VL04 of the \texttt{ITU-T P.1203} collection, it agrees with
subjective scores more closely than the standardized model on the same
sessions, at a Krippendorff alpha of 0.974 over repeated scoring. As the
anchor, it places four studies with no shared content, no shared viewers and
four different scales on one common scale, with a held-out error between 7.8
and 13.3 points out of 100.
Five limitations bound the result. The session report does not represent
conditions that differ only in encoding quality at a fixed resolution. Exact
numerical repetition is not attained on hosted inference. The rubric wording
is part of the instrument. The rater carries a device offset on the UHD
television. Studies with fewer than 60 sessions require a straight line
instead of the logistic curve. Future work concerns longer sessions, modern
codecs, and the harmonization of ground truths in QoE-based anomaly
detection.

\raggedbottom
\bibliographystyle{IEEEtran}
\bibliography{references}

\begingroup

\begin{IEEEbiography}[{\includegraphics[width=1in,height=1.25in,clip,keepaspectratio]{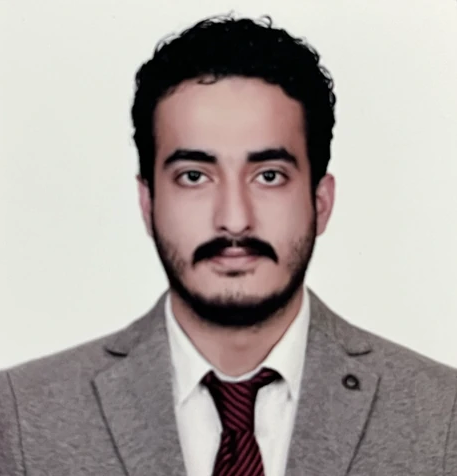}}]{Parsa Hassani Shariat Panahi}
received his B.Sc. degree in Computer Engineering from Azad University South Tehran Branch, Tehran, Iran, and the M.Sc. degree in Computer Engineering - Computer Networks from Iran University of Science and Technology, Tehran, Iran. His research interests include cellular networks, QoE assessment, telecommunication networks, wireless communication, and machine learning. He can be reached at parsa\_hassani@comp.iust.ac.ir.
\end{IEEEbiography}
\vfill
\vspace{-4em}
\begin{IEEEbiography}[{\includegraphics[width=1in,height=1.25in,clip,keepaspectratio]{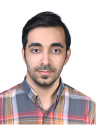}}]{Amir Hossein Jalilvand}
received his B.Sc. degree in Computer Engineering from Bu-Ali Sina University, Hamadan, Iran, and the M.Sc. degree in Computer Engineering - Computer Architecture from Iran University of Science and Technology, Tehran, Iran. He is currently pursuing his Ph.D. in Computer Engineering. His research interests include cellular networks, stochastic and unary computing, computer architecture, fuzzy logic, and machine learning. He can be reached at jalilvand\_a@comp.iust.ac.ir.
\end{IEEEbiography}

\vfill
\vspace{-4em}
\begin{IEEEbiography}[{\includegraphics[width=1in,height=1.25in,clip,keepaspectratio]{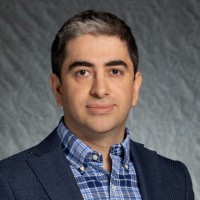}}]{M. Hassan Najafi}
received his Ph.D. in electrical and electronics engineering from the University of Minnesota-Twin Cities, Minneapolis, MN, USA, in 2018. He is currently an Associate Professor at the Electrical, Computer, and Systems Engineering Department at Case Western Reserve University. His research interests include stochastic and approximate computing, unary processing, in-memory computing, and hyperdimensional computing. He has authored/coauthored more than 120 peer-reviewed papers and has been granted 12 U.S. patents. Dr. Najafi received the NSF CAREER Award in 2024, the Best Paper Award at GLSVLSI'23 and ICCD'17, and the 2018 EDAA Outstanding Dissertation Award. He is a senior member of IEEE and of the U.S. National Academy of Inventors. He can be reached at najafi@case.edu.
\end{IEEEbiography}
\endgroup

\end{document}